\documentclass[12pt]{article}
\usepackage[dvips]{color}
\usepackage{epsfig}
\usepackage{graphicx}
\begin{document}
\title {Validity of Born Approximation for Nuclear Scattering in Path Integral
Representation}
\author{M. R. Pahlavani $^{1,}$\thanks{Email:
m.pahlavani@umz.ac.ir}\hspace{1mm}, R. Morad $^{1,}$\thanks{Email:
r.morad@umz.ac.ir}\hspace{1mm} \\
$^1$ {\small {\em  Department of Physics, Faculty of science, Mazandaran University,}}\\
{\small {\em P .O .Box 47415-416, Babolsar, Iran}}\\
}
\maketitle
\begin{abstract}
\noindent The first and second Born approximation are studied with
the path integral representation for $ {\cal T} $ matrix. The $
{\cal T}$ matrix  is calculated for Woods-Saxon potential
scattering. To make corresponding integrals solvable analytically,
an approximate function for the Woods-Saxon potential is used.
Finally it shown that the Born series is converge at high energies
and orders higher than two in Born
approximation series can be neglected.
\end{abstract}
\section{INTRODUCTION }
There is a general idea that there are several ways to describe
nature. In spite of equality of these ways, they are different in
prediction of new laws of physics. Before testing the power of the
predictability of a new formalism, in the first step, it is
necessary to test it with other laws of physics. We are all familiar
with the standard formulations of quantum mechanics, developed more
or less concurrently by Schroedinger, Heisenberg and others in the
1920's, and shown to be equivalent to one another soon thereafter.
In 1933, Dirac made the observation that the action plays a central
role in classical mechanics (he considered the Lagrangian
formulation of classical mechanics to be more fundamental than the
Hamiltonian one), but that it seemed to have no important role in
quantum mechanics as it was known at the time. He speculated on how
this situation might be rectified, and he arrived at the conclusion
that (in more modern language) the propagator in quantum mechanics
"corresponds to" $exp(iS/\hbar)$, where $S$ is the classical action
evaluated along the classical path. In 1948, Feynman developed
Dirac's suggestion, and succeeded in deriving a third formulation of
quantum mechanics, based on the fact that the propagator can be
written as a sum over all possible paths (not just the classical
one) between the initial and final points. Each path contributes
$exp(iS/\hbar)$ to the propagator. So while Dirac considered only
the classical path, Feynman showed that all paths contribute: in a
sense, the quantum particle takes all paths, and the amplitudes for
each path add according to the usual quantum mechanical rule for
combining amplitudes[1, 2]. Path integrals give us no dramatic new
results in the quantum mechanics of a single particle. Indeed, most
if not all calculations in quantum mechanics which can be done by
path integrals can be done with considerably greater ease using the
standard formulations of quantum mechanics. path integrals turn out
to be considerably more useful in more complicated situations, such
as field theory. But even if this were not the case, it is believed
today that path integrals would be a very worthwhile contribution to
our understanding of quantum mechanics. Firstly, they provide a
physically extremely appealing and intuitive way of viewing quantum
mechanics: anyone who can understand Young's double slit experiment
in optics should be able to understand the underlying ideas behind
path integrals. Secondly, the classical limit of quantum mechanics
can be understood in a particularly clean way via path integrals. It
is in quantum field theory, both relativistic and nonrelativistic,
that path integrals (functional integrals is a more accurate term)
play a much more important role, for several reasons. They provide a
relatively easy road to quantization and to expressions for Green's
functions[3, 4, 5], which are closely related to amplitudes for
physical processes such as scattering and decays of particles.
Furthermore, the close relation between statistical mechanics and
quantum mechanics, or statistical field theory and quantum field
theory, is plainly visible via path integrals. Now, path-integral
formalism is widely used in many branches of theoretical physics and
in particular in nuclear physics[6]. The existence of an strong
nuclear force in atomic nuclei revealed by the exceptional role of
the nuclear magic number provides the foundation of the nuclear
shell model. This strong force is believed to be approximated most
closely by a Woods-Saxon potential [7] either from analyzing the
radial dependence of the nuclear central force or by deriving it
from a microscopic two body force acted in neutron proton scattering
[8]. The spherical Woods-Saxon potential that was used as a major
part of nuclear shell model, was successful to deduce the nuclear
energy levels [9]. Also it was used as central part for the
interaction of neutron with heavy nucleus [10]. The Woods-Saxon
potential was used as a part of optical model in elastic scattering
of some ions with heavy target in low range of energies [11]. In
this paper we will use the path integral representation of the $
{\cal T}$ matrix in potential scattering for driving the different
orders of Born approximation with the Woods-Saxon potential. This
representation is not a phase-space path integral but it is a
particular path integral over velocities that reduced the complexity
of path-integral[12]. In the limit of large scattering times where
energy conservation, the "dangerous" phases from $ {\cal S} $ matrix
are created. The "Phantom" degrees of freedom is utilized to
eliminate these phases. In addition, energy conservation is applied
by imposing a Faddeev-Popov-like constraint in the velocity path
integral[12]. In Sec.II we outline the $ {\cal T} $ matrix
representation from the velocity path integrals. In order to obtain
different terms of Born approximation, we use path integral
representation for the scattering in presence of Woods-Saxon
potential in Sec.III. In this investigation we show that the Born
approximation is a converge series, therefore in elastic scattering
analysis, we can neglect the orders higher than two in Born series.
\section{Path Integrals for the $ {\cal T}$ matrix }
In the framework of nonrelativistic potential scattering, consider a
central potential, $V(r)$ that vanishes at infinity.
$\textbf{k}_{i}$ and $\textbf{k}_{f}$ are the initial and final
momentum of a particle with mass $m$ (see Figure.1). In this
calculation, supposed the scattering states are normalized and
$\hbar = 1 $[12], so
\begin{equation}
\langle\varphi_{f}|\varphi_{i}\rangle=(2\pi)^{3}\delta^{3}(k_{i}-k_{f}).
\label{ec1}
\end{equation}
The ${\cal S}$ matrix is the matrix element of the evolution
operator in the interaction picture that taken between scattering
states and calculated at asymptotic times:
\begin{equation}
{\cal S}_{i \longrightarrow f} = \lim_{T  \longrightarrow \infty} \>
\left < \phi_f \left | \, \hat U_I (T,-T) \, \right | \phi_i \right
> = \lim_{T  \longrightarrow \infty} \> e^{i (E_i + E_f) T} \, \left < \phi_f
\left | \, \hat U(T,-T) \, \right | \phi_i \right > , \label{ec2}
\end{equation}
where the $\hat U_I(T,-T)$ is the time evolution operator that is
defined as follows
\begin{equation}
 \hat U_I(t_b,t_a) = e^{i\hat H_0 \, t_b} \, \exp \left [ - i
\hat H \, (t_b - t_a ) \, \right ] \, e^{-i\hat H_0 \, t_a} \> \>.
\label{ec3}
\end{equation}
\begin{tabular*}{2cm}{cc}
\hspace{3cm}\includegraphics[scale=0.7]{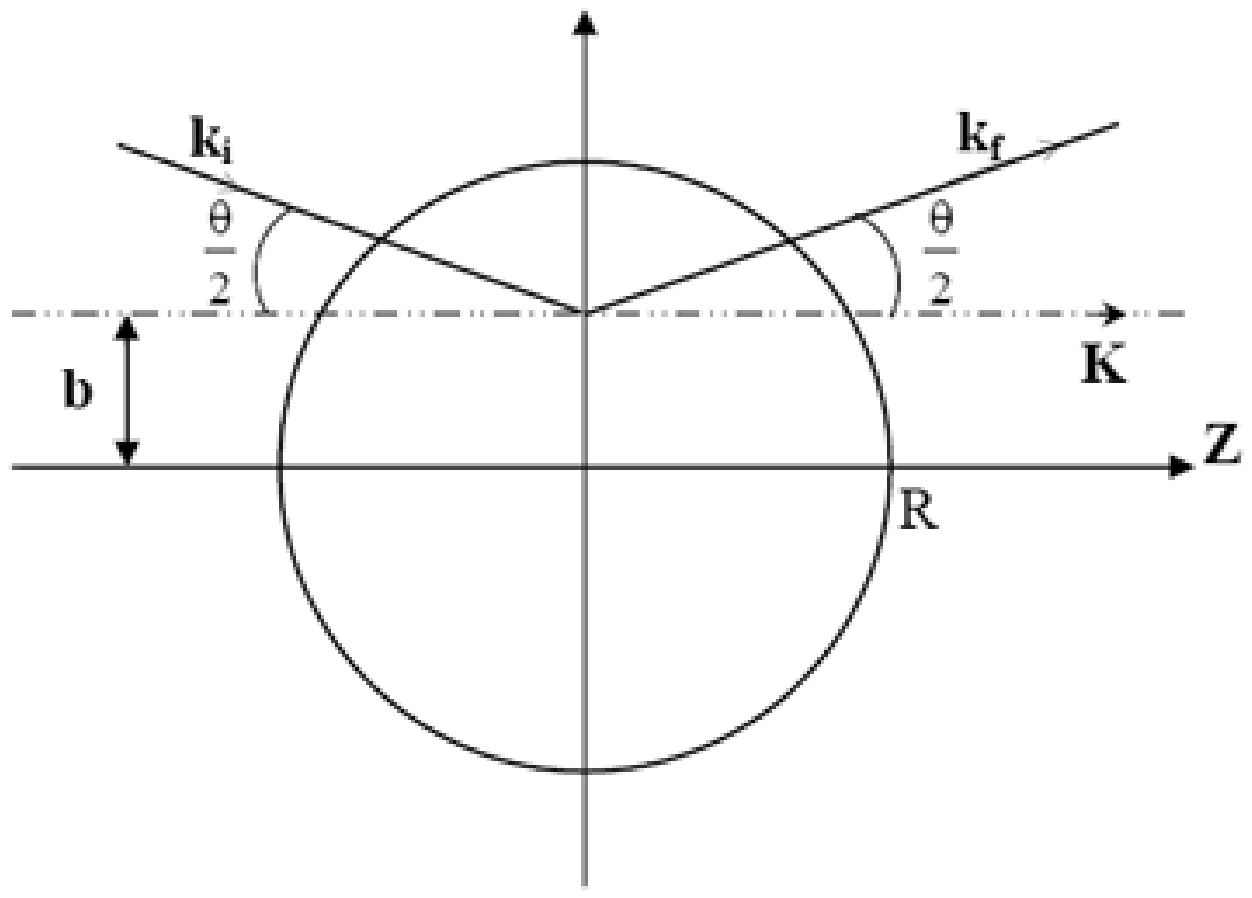}\\
\end{tabular*}\\
Figure1: Scattering geometry for a potential of radius $R$, the
impact parameter $b$. Incoming and outgoing momenta are $ \> {\bf
k}_{i,f} \> $, and the mean momentum is  $ \textbf{K} =
(\textbf{k}_i +
\textbf{k}_f)/2 $.\\\\
The ${\cal T}$ matrix  is defined usually by subtract the identity
from ${\cal S}$ matrix and factor out an energy conserving Dirac
delta function
\begin{eqnarray}
{\cal S}_{i  \longrightarrow f} = (2 \pi)^3 \delta^{(3)} \left (
\textbf{k}_i -\textbf{k}_f \right )  - 2 \pi i \delta\left ( E_i -
E_f \right ) \, {\cal T}_{i \longrightarrow f} \> , \label{ec4}
\end{eqnarray}
where $ E_i = E_f = E = k^2/(2m)$ is the common scattering energy.
By integrating functionally over velocities instead of
paths[14,15,16] the following formula is achieved[13]
\begin{eqnarray}
\left ( {\cal S} - 1 \right )_{i \longrightarrow f} = \lim_{T
\longrightarrow \infty} \exp \left ( \, i \frac{\textbf{q}^2}{4 m} T
\, \right ) \> \int d^3 r \> e^{- i \textbf{q} \cdot \textbf{r}} \>
\, {\cal N}^{\, 3}(T,-T) \> \int {\cal D}^3 v \> \exp \left [ \, i
\int_{-T}^{+T} dt \, \frac{m}{2} \textbf{v}^2(t) \, \right ]
\nonumber \\
\hspace{4cm} \times \left \{ \, \exp \left [ \> - i \int_{-T}^{T} dt
\> V \left ( \, \textbf{r} + \frac{\textbf{K}}{m} t +
\textbf{x}_v(t) \right ) \right ] - 1 \, \right \}\label{ec5}
\end{eqnarray}
where
\begin{eqnarray}
{\cal N}(t_a, t_b) :=  \left ( \> \int {\cal D}v \> \exp \left [ \,
i \int_{t_a}^{t_b} dt \>\frac{m}{2} v^2 (t) \, \right ] \> \right
)^{-1}, \label{ec6}
\end{eqnarray}
and
\begin{eqnarray}
\textbf{x}_v(t) = \frac{1}{2} \int_{-T}^{+T} dt'  {\rm sgn}(t-t') \,
\textbf{v}(t') \> . \label{ec7}
\end{eqnarray}
The momentum transfer and the mean momentum are defined by
\begin{eqnarray}
\textbf{q} = \textbf{k}_f - \textbf{k}_i \> , \hspace{0.3cm}
\textbf{K} = \frac{1}{2} \left ( \textbf{k}_i + \textbf{k}_f \right
) \> . \label{ec8}
\end{eqnarray}
It can be shown that the 'dangerous phases' $q^{2}T/4m $ that are
proportional to T, are canceled in each order of perturbation theory
by introducing 'Phantom' degrees of freedom( dynamical variables
with the wrong sight kinetic term)[12], so that the limit $ T \to
\infty $ can indeed be applied. Then the following path-integral
representation for the ${\cal S}$ matrix is obtained[12],
\begin{eqnarray}
\left ( {\cal S} - 1 \right )_{i \longrightarrow f} = \lim_{T
\longrightarrow \infty} \> \int d^3 r \> e^{- i \textbf{q} \cdot
\textbf{r}} \> \left | {\cal N} (T,-T) \right|^6 \, \int {\cal D}^3
v \, {\cal D}^3 w  \> \exp \left [ \, i \int_{-T}^{+T} dt \,
\frac{m}{2} \left ( \textbf{v}^2(t) -\textbf{w}^2(t) \right ) \,
\right ]   \nonumber \\
\cdot \left \{ \, \exp \left [ \> - i \int_{-T}^{+T} dt \> V \left (
\, \textbf{r} + \frac{\textbf{K}}{m} t + \textbf{x}_v(t) -
\textbf{x}_w(0) \right ) \right ] - 1 \, \right \} \> , \label{ec9}
\end{eqnarray}
where $\textbf{w}(t)$ is a three-dimensional 'antivelocity'. This is
similar to the Lee-Wick approach to quantum Electrodynamics where
they introduced the fields with a wrong sign kinetic term to remove
all infinities[17,18,19]. To extract the ${\cal T} $ matrix from the
${\cal S}$ matrix for weak interaction, we can develop in powers of
the potential and factor out in each order an energy conserving
$\delta $ function. But to achieve this without a perturbative
expansion of ${\cal S}$ matrix usually the trick which Faddeev and
Popov have introduced in field theory for quantization of
non-Abelian gauge theories is used [20] and the following expression
for the $ {\cal T} $ matrix  is obtained[12],
\begin{eqnarray}
{\cal T}_{i \longrightarrow f}^{(3-3)} = i \frac{K}{m} \> \int d^2 b
\> e^{- i \textbf{q} \cdot \textbf{b} } \> \left | {\cal N} \right
|^6 \ \int {\cal D}^3 v \, {\cal D}^3 w \> \exp \left \{ \, i
\int\limits_{-\infty}^{+\infty} dt \,  \frac{m}{2} \left [
\textbf{v}^2(t) -\textbf{w}^2(t) \right \} \, \right ] \,
\> \\
\nonumber \times \Biggl \{ e^{ i \chi_{\textbf{K}}(
\textbf{b},\textbf{v},\textbf{w})} \, - \, 1 \> \Biggr \} \> .
\label{ec10}
\end{eqnarray}
Here the limit $T \to \infty $  is taken and the corresponding
Gaussian normalization factor is written as
\begin{eqnarray}
{\cal N} \> := \> {\cal N}(+\infty,-\infty) \> .\label{ec11}
\end{eqnarray}
In this equation the phase $ \chi_{\textbf{K}}$ is defined as
\begin{eqnarray}
\chi_{\textbf{K}}(\textbf{b},\textbf{v},\textbf{w}) = -
\int_{-\infty}^{+\infty} dt \> V \left ( \textbf{b} +
\frac{\textbf{K}}{m} \, t  + \textbf{x}_v (t) - \textbf{x}_w(0) -
\lambda \hat \textbf{K} \right ) , \label{ec12}
\end{eqnarray}
where $\textbf{b}$ is the impact parameter. $ \theta$ is the
scattering angle,one may obtained
\begin{eqnarray}
q \equiv | \textbf{q}| = 2 k \sin \left (\frac{\theta}{2} \right )
\> , \hspace{0.3cm} K \> \equiv \> | \textbf{K}| =  k \cos \left
(\frac{\theta}{2} \right ) \> . \label{ec13}
\end{eqnarray}
For $t_0=0$  that is the most symmetric choice the gauge parameter,
$ \lambda = K t_0/m $, is zero. The superscript "3-3" indicates that
there are three-dimensional antivelocity that used to cancel
divergent phases in the limit of asymptotic times moreover the
three- dimensional velocity variables. Then by expanding the
exponent in powers of the potential the complete Born series is
reproduced from this path-Integral representation as follows[12],
\begin{eqnarray}
{\cal T}_{i \longrightarrow f} \> =: \> \sum_{n=1}^{\infty} {\cal
T}_n , \label{ec14}
\end{eqnarray}
with
\begin{eqnarray}
{\cal T}_n^{(3-3)} = i \frac{K}{m} \, \frac{(-i)^n}{n!} \int d^2b \,
e^{-i \textbf{q} \cdot \textbf{b}} \> \prod_{i=1}^n \left (
\int_{-T}^{+T} dt_i \int \frac{d^3p_i}{(2 \pi)^3} \, \tilde
V(\textbf{p}_i) \right ) \\
\nonumber  \times \exp \left \{ i
\sum_{i=1}^n \textbf{p}_i \cdot \left [ \textbf{b} + \textbf{x}_{\rm
ref}(t_i) \right ] \right \} \, G_n^{(3-3)} \> . \label{ec15}
\end{eqnarray}
Where $G_n^{(3-3)}$ is evaluated as
\begin{eqnarray}
G_n^{(3-3)} = \exp \left \{ \> -\frac{i}{4m} \sum_{i,j=1}^n
\textbf{p}_i \cdot \textbf{p}_j \, \left ( \, T - |t_i - t_j| - T \,
\right ) \> \right \} \> , \label{ec16}
\end{eqnarray}
and
\begin{eqnarray}
\textbf{x}_{\rm ref}(t)  = \frac{\textbf{K}}{m} \, t .\label{ec17}
\end{eqnarray}
$\tilde V(\textbf{p}_i)$ is the fourier transform of potential
\begin{eqnarray}
\tilde V(\textbf{p}_i) = \int d^3r  \, V(r) \, e^{i \textbf{q} \cdot
\textbf{r}} \>. \label{ec18}
\end{eqnarray}
{\centering\section{Born Approximation \label{sec.III} }}
In this
section, we calculated the first and the second order of Born
approximation from the path integral formula for ${\cal T}$ matrix.
For first Born approximation from the Eq.(15) we have
\begin{eqnarray}
{\cal T}_1^{(3-3)} = \frac{K}{m} \, \int d^2b \, e^{-i \textbf{q}
\cdot \textbf{b}} \int_{-T}^{+T} dt \int \frac{d^3p}{(2 \pi)^3} \,
\tilde V(\textbf{p}) \times \exp \left \{ i \textbf{p} \cdot \left (
\textbf{b} + \frac{\textbf{K}}{m} \, t \right ) \right \} \, .
\label{ec19}
\end{eqnarray}
That with integration over b and t, we obtain
\begin{eqnarray}
{\cal T}_1^{(3-3)} = \tilde V( \textbf{p}_\bot=\textbf{q} ,
\textbf{p}_\|=0 )\, = \, \tilde V(\textbf{q}). \label{ec20}
\end{eqnarray}
Where $ \textbf{p}_\perp $ and $\textbf{p}_\parallel$ are the
components of $ \textbf{p}$ perpendicular and parallel to
$\textbf{K}$ respectively. Also $\textbf{q}$ is a vector in the
plane which is perpendicular to $\textbf{K}$. The second Born
approximation is calculated in the following formula
\begin{eqnarray}
{\cal T}_2^{(3-3)} = \frac{-i K \pi}{m} \, \int \int d^3p_1 d^3p_2
\tilde V(\textbf{p}_1) \tilde V(\textbf{p}_2)  \delta^{(2)} \left
(\textbf{q}- \textbf{p}_{1\perp} - \textbf{p}_{2\perp} \right
)\\
\nonumber \times \delta \left (\textbf{p}_1 \cdot \frac{K}{m} +
\frac{\textbf{p}_1 \cdot \textbf{p}_2}{2m} \right ) \delta \left (
\textbf{p}_2 \cdot \frac{K}{m} - \frac{\textbf{p}_1 \cdot
\textbf{p}_2}{2m} \right ). \label{ec21}
\end{eqnarray}
To simplify the above equation, we suppose that $\textbf{p}_1 \cdot
\textbf{p}_2 = 0 $, then the Eq.(21) can be rewritten as
\begin{eqnarray}
{\cal T}_2^{(3-3)} = \frac{-i m \pi}{K} \, \int d^2p \, \tilde
V(\textbf{p}) \, \tilde V(\textbf{q}-\textbf{p}). \label{ec22}
\end{eqnarray}
Integration over $\textbf{p}$ can been done in the plane
perpendicular to $\textbf{K}$.
Now, we apply this formalism to
calculate the first and second Born approximation for Woods-Saxon
potential.
The Woods-Saxon potential can be defined by
\begin{eqnarray}
V(r) = \frac{V_0}{1 +\exp(\frac{r-R_0}{a})}.
\label{ec23}
\end{eqnarray}
This potential may be represented, with precision better than $3 \%$
for any r value ( see figure.2 ) by
\begin{eqnarray}
V(r) = \frac{V_0}{1 +\exp(\frac{r-R_0}{a})} = \,  V_0 \,
C(\frac{r-R_0}{a})\label{ec24}
\end{eqnarray}
\begin{eqnarray}
C(x\leq 0) = 1\,-\,\frac{7}{8}\, e^x \,+ \, \frac{3}{8}\, e^{2x}
\label{ec25}
\end{eqnarray}
\begin{eqnarray}
C(x\geq 0) = e^{-x}\, ( \, 1\,-\,\frac{7}{8}\, e^{-x} \,+ \,
\frac{3}{8}\, e^{-2x} ) . \label{ec26}
\end{eqnarray}
\begin{figure}[bth]
\centerline{\includegraphics[width=10cm]{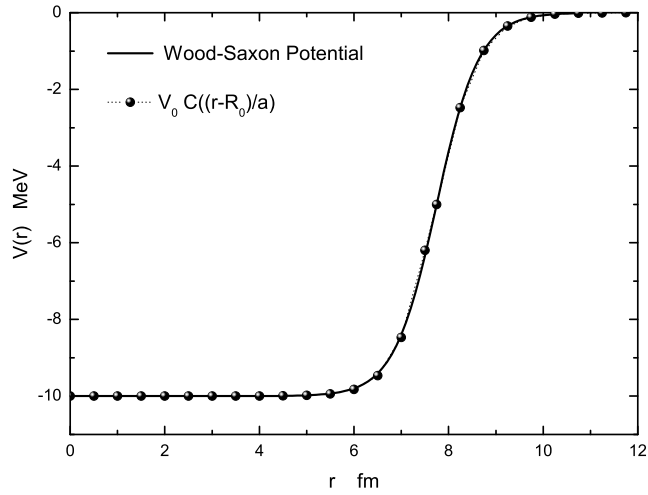}} \caption{Nuclear
potential for the $ ^{16}O+^{20}Ne $ scattering represented by
Woods-Saxon potential(solid line) and the approximate function $ V_0
\, C((r-R_0)/a) $ (symbol $\bullet$). \label{fig2}}
\end{figure}
This approximation is particularly useful in obtaining analytical
expression for integrals that involve the Woods-Saxon potential. A
such approximation is used for tow parameter Fermi distribution
density function[21]. With this approximation the fourier transform
of Woods-Saxon potential is
\begin{eqnarray}
\tilde V(\textbf{q}) = \pi\, V_0 \, a^3\,[ \,
\frac{7e^{-R_0/a}}{(q^2a^2+1)^2} \, -
\,\frac{6e^{-2R_0/a}}{(q^2a^2+4)^2} \,] , \label{ec27}
\end{eqnarray}
where q is the magnitude of $\textbf{q}$. And the second Born
approximation can be written as,
\begin{eqnarray}
{\cal T}_2^{(3-3)} = \frac{m \pi^3 V_0^2 a^6}{K}\, [\, 49\,
e^{-2R_0/a} \, g_1(q)\, -\, 42\, e^{-3R_0/a} \,g_2(q) \, +\, 36\,
e^{-4R_0/a} \, g_3(q)\,],
\label{ec28}
\end{eqnarray}
where $g_1(q),g_2(q)$ and $g_3(q)$ are solvable integrals over p,
but their results are too complicated, thus their results are not
appeared here. It should be noted that q is a function of scattering
energy (E) and angle($\theta$), therefore
\begin{eqnarray}
g_1(q) =  \int d p \, \frac{p}{(p^2a^2+1)^2((q-p)^2a^2+1)^2},
\label{ec29}
\end{eqnarray}
\begin{eqnarray}
g_2(q) =  \int d p \, p \,[\,
\frac{1}{(p^2a^2+1)^2\,((q-p)^2a^2+4)^2} \,+\,
\frac{1}{(p^2a^2+4)^2\,((q-p)^2a^2+1)^2}\,], \label{ec30}
\end{eqnarray}
\begin{eqnarray}
g_3(q) =  \int d p \, \frac{p}{(p^2a^2+4)^2((q-p)^2a^2+4)^2}.
\label{ec31}
\end{eqnarray}
We wish to calculate the first and second Born approximation for the
$^{16}O\,+\,^{20}Ne$. The parameters of Woods-Saxon potential for
this system are presented in table.1. The first and second Born
approximation, for this reaction, as a function of scattering energy
and angle are shown in Figure.3 and Figure.4 respectively. The
graphs show that the values of $T_1$ and $T_2$ are significant in
small angles and extremely reduced with increasing of angle. For
comparison between first and second Born approximation, we show
these approximation versus the scattering angle $\theta$, at E=24.5
MeV in Figure.5. It can be seen from this Figure that the value of
$T_2$ is smaller than $T_1$ and can be neglected. Also in fig.6 the
values of $T_1$ and $T_2$ for $\theta=0^\circ , \theta=45^\circ,
\theta=90^\circ$ versus scattering energy are plotted. These graphs
show that the value of $T_2$, in all angles and in energies below
and near the coulomb barrier, is very smaller than $T_1$. This is
arises from fact that the Woods-Saxon is a short ranged potential.
In low energies the repulsive coulomb force prevents the nuclei to
be close. So it is not convenient to use the nuclear potential to
study the scattering. In this energies, the scattering is Ratherford
scattering. Finally we can conclude that the Born approximation is a
converge series and we can neglect higher order of Born
approximation in elastic scattering studies.
\begin{table}[tbh]
\caption{Woods-Saxon parameters for $^{16}O\,+\,^{20}Ne$
reaction.($R_0=r_0(A_1^{1/3}+A_2^{2/3}$))}[22]
\begin{center}
\begin{tabular}{cccc}
  \hline\hline
  $ System $ & $V_0 (MeV)$  & $ r_0  (fm)$ & $a  (fm) $ \\
  \hline\hline
  $^{16}O+^{20}Ne  $ & 10  & 1.48 & 0.45 \\
     \hline
\end{tabular}
\label{tab1}
\end{center}
\end{table}
\begin{figure}[bth]
\centerline{\includegraphics[width=15cm]{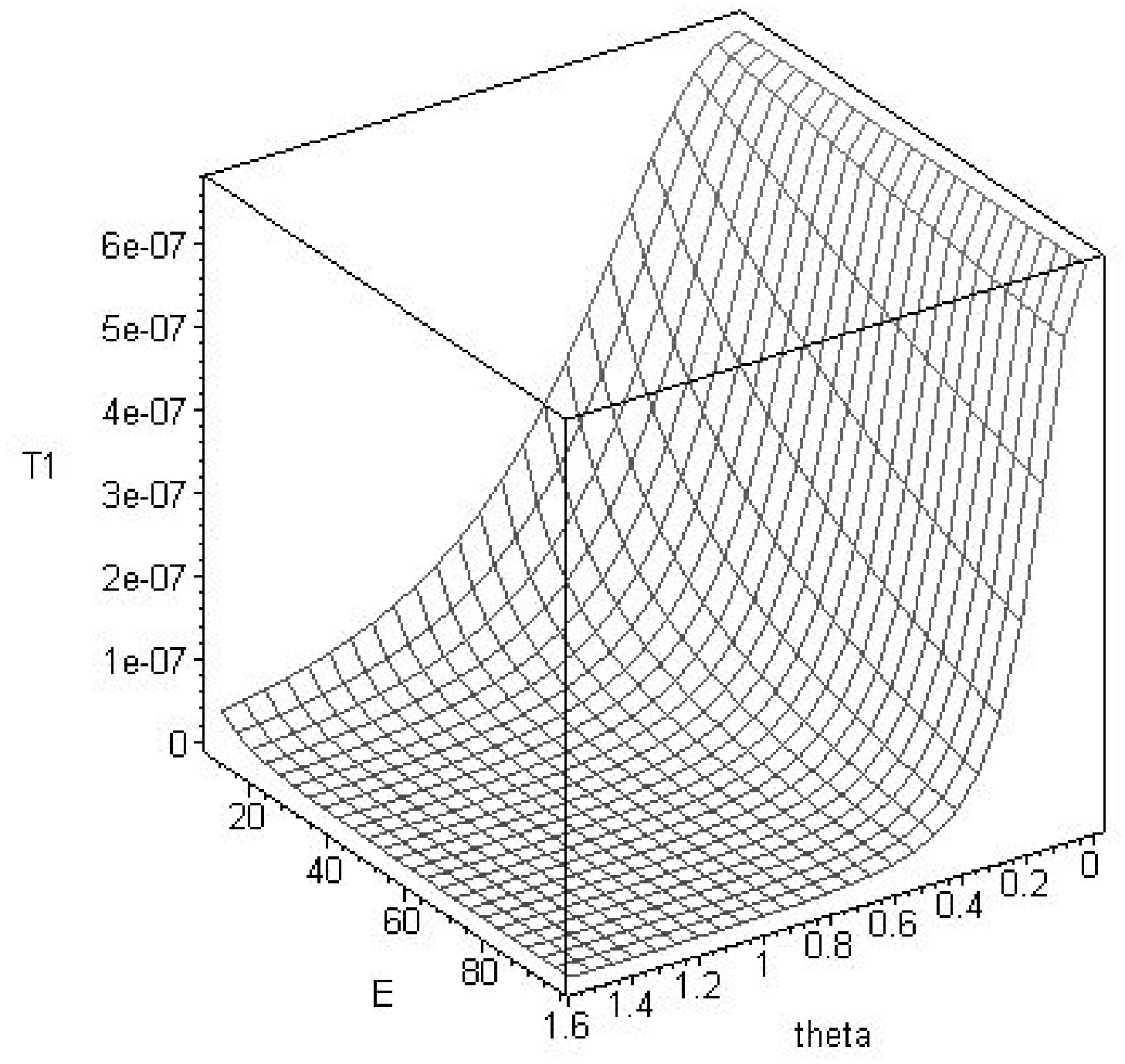}} \caption{ First
Born approximation versus scattering energy and angle($\theta$) for
$ ^{16}O+^{20}Ne $.  \label{fig3}}
\end{figure}
\begin{figure}[bth]
\centerline{\includegraphics[width=15cm]{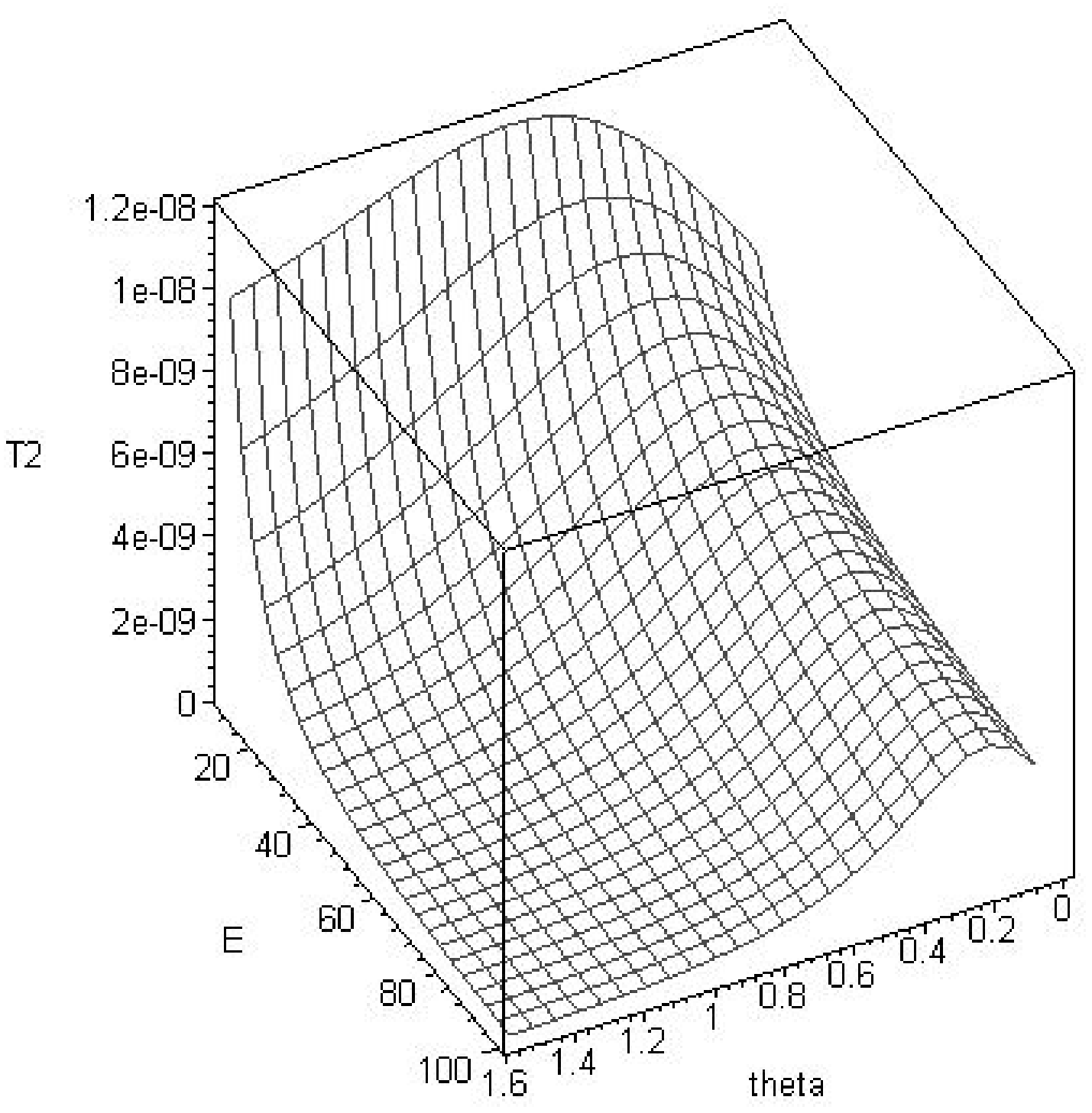}} \caption{Second
Born approximation versus scattering energy and angle($\theta$) for
$ ^{16}O+^{20}Ne $ \label{fig4}}
\end{figure}
\begin{figure}[bth]
\centerline{\includegraphics[width=15cm]{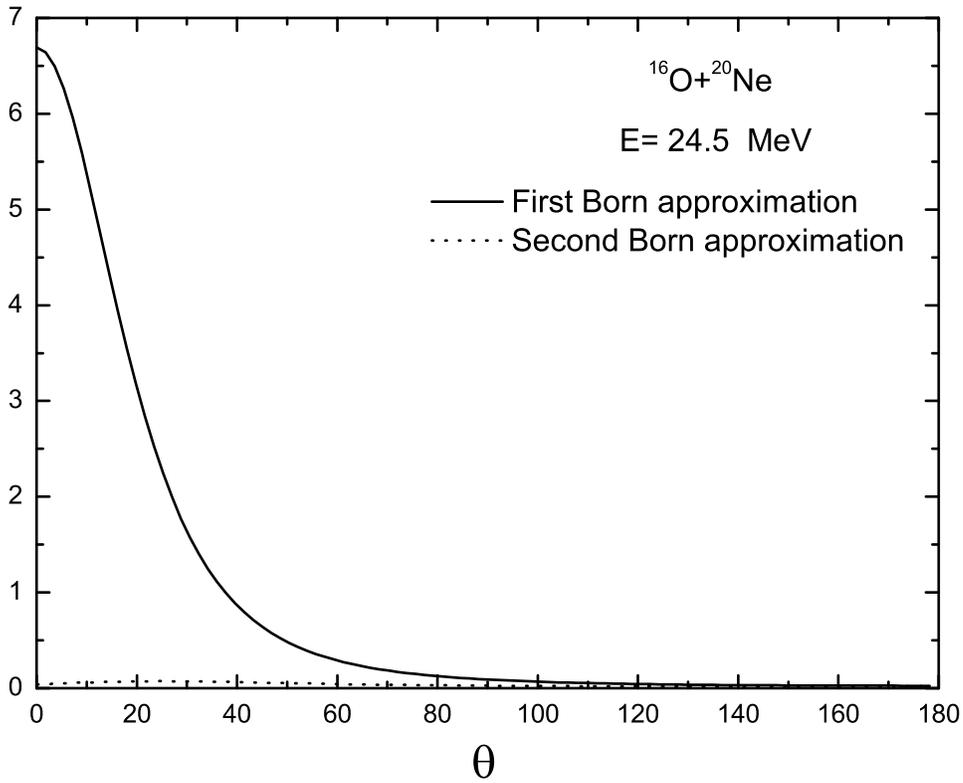}} \caption{The
first and second Born approximation calculated with path integral
formalism for $^{16}O\,+\,^{20}Ne $ scattering versus $\theta$ in E
= 24.5 MeV. solid line shows the first Born approximation and dash
line shows the second Born approximation. ( in scale $10^{-7}$).
\label{fig5}}
\end{figure}
\begin{figure}[bth]
\centerline{\includegraphics[width=22cm]{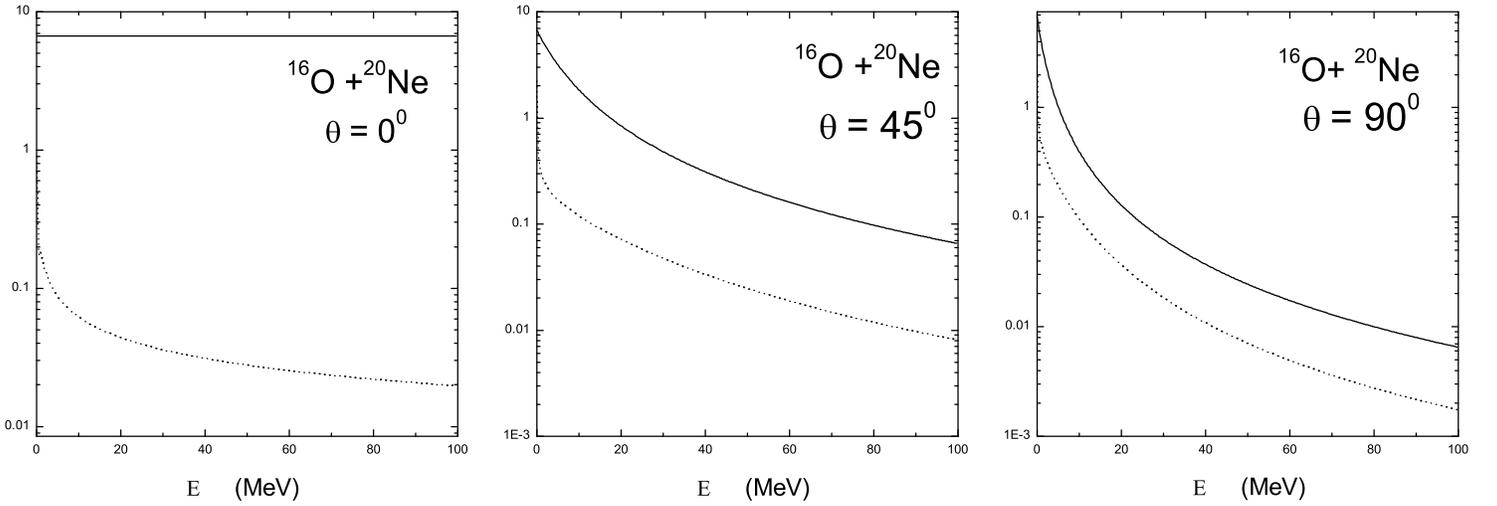}} \caption{The
first and second Born approximation calculated with path integral
formalism for $^{16}O\,+\,^{20}Ne $ scattering versus scattering
energy (E) in three angle : $\theta=0^\circ$, $\theta=45^\circ$ and
$\theta=90^\circ$. solid line shows the first Born approximation and
dash line shows the second Born approximation. ( in scale
$10^{-7}$). \label{fig6}}
\end{figure}

\end{document}